\documentclass{iau}

\PassOptionsToPackage{pdftex}{hyperref}
\usepackage{amsmath}
\usepackage{graphicx}
\usepackage{multirow}

\makeatletter
\renewcommand{\doitext}{}
\let\@doi\@empty
\makeatother

\lefttitle{Shin et al.}
\righttitle{Discovery of GeV--TeV emission from UKS 1}

\jnlPage{1}{4}
\jnlDoiYr{2025}

\aopheadtitle{Proceedings IAU Symposium 398}

\title{Second Discovery of GeV--TeV Connection from the Globular Cluster UKS 1}

\author{
  Jiwon Shin$^1$, 
  C.~Y. Hui$^2$, 
  Sangin Kim$^1$, 
  Kwangmin Oh$^3$, 
  Ellis R. Owen$^{4,5}$
}

\affiliation{
$^1$ Department of Earth, Environmental \& Space Sciences, Chungnam National University, Daejeon 34134, Republic of Korea \\
$^2$ Department of Astronomy and Space Science, Chungnam National University, Daejeon 34134, Republic of Korea \\
$^3$ Department of Physics and Astronomy, Michigan State University, East Lansing, MI 48824, USA \\
$^4$ Astrophysical Big Bang Laboratory (ABBL), RIKEN Pioneering Research Institute (PRI), Wak\=o, Saitama 351-0198, Japan \\
$^5$ Theoretical Astrophysics, Department of Earth and Space Science, Graduate School of Science, The University of Osaka, Toyonaka, Osaka 560-0043, Japan
}

\begin{document}

\begin{abstract}

Using 16 years of data collected by {\it Fermi} Large Area Telescope and 1523 days of survey data from High Altitude Water Cherenkov (HAWC) Observatory, we discovered the long-sought second GeV–TeV connection towards the globular cluster (GC) UKS 1 \citep{Shin2025A&A} . Gamma-ray spectroscopy suggests that the GeV emission can be attributed to both the pulsar magnetosphere and inverse Compton scattering (ICS) by the pulsar wind. In particular, the TeV peak is displaced from the cluster center by several tidal radii in the trailing direction of the GC’s proper motion. This alignment supports a scenario in which relativistic leptons, likely driven by a millisecond pulsar population, produce very-high-energy (VHE) gamma rays via ICS within a bow shock tail. Our findings not only highlight GCs as potential sources of VHE gamma rays, but also offers a rare opportunity to probe cosmic ray transport in the Milky Way by studying particle propagation and anisotropic gamma-ray production associated with the extended, offset TeV feature of UKS 1.
\end{abstract}


\maketitle

\section{Introduction}
Owing to frequent stellar encounter, compact binaries can be formed dynamically inside a globular cluster (GC) including millisecond pulsars \citep[MSPs;][]{Hui2010.714}. Since the magnetosphere of MSPs can accelerate electrons/positrons to relativistic speeds with Lorentz factor $>10^{6}$, their collective contributions make GCs as $\gamma$-ray emitters. Shortly after {\it Fermi} Large Area Telescope (LAT) was launched, a number of GCs were detected in the GeV band, forming a distinct source class \citep{Tam2016JASS,GC_LAT,4fgldr4}. 

There are two possible origins of the GeV emission from a GC. Since a number of $\gamma-$ray GCs have spectra which resemble that of a MSP, namely a power-law with an exponential cutoff at energies of $\lesssim10$~GeV, some studies argued that the GeV emission from GCs entirely come from the collective pulsed emission originating within the magnetosphere \citep[e.g.][]{GC_LAT}. On the other hand, some spectra of GCs show hard tails at energies larger than few tens of GeV \citep{Song_2021}. We have shown that this is consistent with a scenario where these $\gamma-$rays originate from the inverse Compton scattering (ICS) between the relativistic pulsar wind and the ambient soft photons \citep{Cheng_2010,Hui_2011}. 

For a GC moving in the rest frame of interstellar medium (ISM) with a speed of order 100 km~s$^{-1}$, a bow shock can be formed. The pulsar wind particles can be further accelerated at the shock front. Through the ICS between these re-accelerated leptons and  soft photons, very-high-energy (VHE) $\gamma-$rays can be produced \citep{Bednarek2007MNRAS, Bednarek_2014}. Possible evidence of this process was first found in the direction of Terzan 5 \citep{HESS2011AnA}. An asymmetric TeV nebula has been detected around this GC with an offset between $\gamma-$ray emission peak and cluster center. In contrast, the GeV emission coincides with Terzan 5 \citep{Terzan5_LAT}.  

 The relativistic particles emanating from the bow shock are expected to be strongly oriented along the trailing magnetotail, with anisotropic ICS emission directed along their propagation vector. Self-scattering leads to a gradual broadening of the particle pitch angle distribution, isotropizing their emission and allowing it to become visible some distance away from the acceleration site \citep{Krumholz_2024}. Under this scenario, TeV nebulae powered by GCs can provide a key to constrain the scattering processes of cosmic rays in the Milky Way.

However, the TeV feature found in the direction of Terzan 5 has remained the only confirmed detection for more than a decade. This has motived our search for possible GeV-TeV GCs in archival data. 

\section{Discovery of GeV-TeV emission from UKS 1}
Using 16 years public data collected by {\it Fermi}-LAT and the 1523 days of 3HWC survey data acquired by High Altitude Water Cherenkov (HAWC) observatory \citep{3hwc}, we reported the detection of a GeV $\gamma$-ray feature at the location of the GC UKS 1, which is associated with a TeV excess identified in the 3HWC survey data \citep{Shin2025A&A}. For details, refer to \citet{Shin2025A&A}. Here we highlight our major results. 

\begin{figure}[t]
 \begin{center}
  \includegraphics[width=1\textwidth]{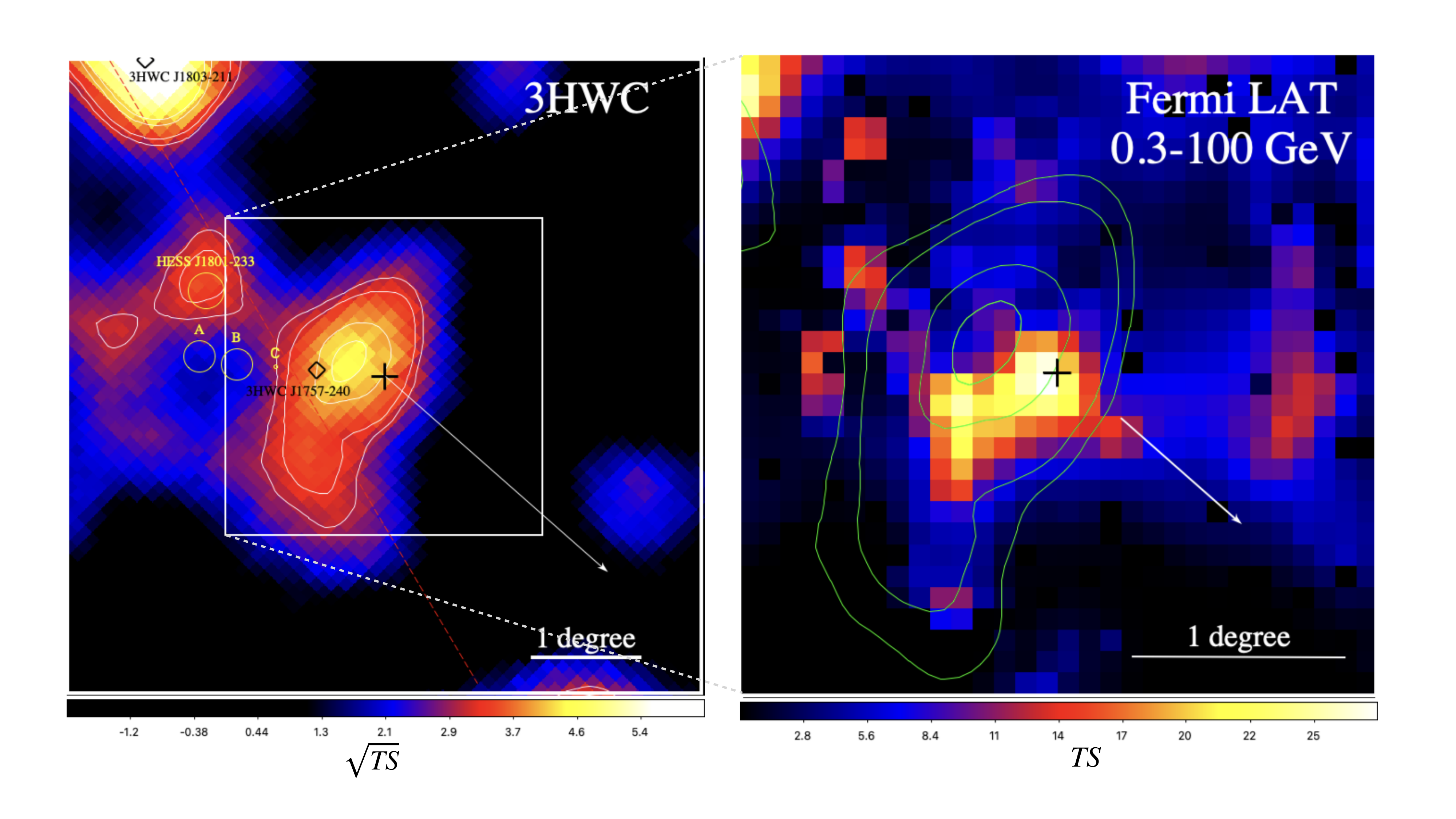}
   \end{center}
  \vspace{-1cm}
  \caption{({\it Left panel:}) $6^{\circ}\times6^{\circ}$ 3HWC significance map around the center of UKS 1 from a point-source search with contours illustrating significance levels of $3\sigma$, $3.3\sigma$, $4\sigma$, and $4.3\sigma$. The location and extent of nearby TeV sources are indicated by yellow circles. The red-dashed lines shows the orientation of Galactic plane. ({\it Right panel:}) Close-up view within $3^{\circ}\times3^{\circ}$ of the {\it Fermi}-LAT Test Statistic (TS) map, with 3HWC contours overlaid. The black crosses and white arrows in both panels illustrate the center and the direction of the proper motion of UKS 1, respectively. Top is north and left is east. A scale bar of $1^{\circ}$ is given in both panels}
  \label{fig:hawc_lat}
\end{figure}

Figure~\autoref{fig:hawc_lat} shows the $\gamma-$ray excess in the direction of UKS 1 as found in the LAT data ({\it right panel}) and the 3HWC data ({\it left panel}), respectively. The TeV excess is found from a point-source search by assuming a power-law energy spectrum with an index of 2.5. The contours illustrate significance levels of $3\sigma$, $3.3\sigma$, $4\sigma$ and $4.3\sigma$. The peak significance of the feature in proximity to UKS 1 is located at 
RA=$17^{\rm h}55^{\rm m}54.38^{\rm s}$, Dec=$-23^{\circ}57^{'}30.96^{''}$ (J2000), with a significance of $4.4\sigma$. It is displaced from the center of UKS 1 by $\sim0.39^\circ$  in the direction of N60.7$^{\circ}$E. Interestingly, we found the displacement almost traces the direction opposite to the proper motion of UKS 1 which is illustrated by the arrow in Figure~\autoref{fig:hawc_lat}. 

To compare the $\gamma$-ray excesses as observed by {\it Fermi} (which is detected at a significance of $\sim6\sigma$) and HAWC, we show a close-up view of the $3^{\circ}\times3^{\circ}$ region centered on UKS 1 as observed by LAT with the 3HWC significance contours overlaid in the right panel of Figure~\ref{fig:hawc_lat}. Both excesses are apparently extended in a similar orientation toward the Galactic plane (the red dashed line in the left panel of Figure~\ref{fig:hawc_lat}). 

The {\it Fermi}-LAT spectrum of UKS is best described a single power-law model with a photon index of $\Gamma=2.3\pm0.5$. There is a clear excess above $\sim10$~GeV. Integrating the model between 0.3-100 GeV yields an energy flux of $f^{\rm LAT}_{\gamma}=(3.24\pm2.74)\times10^{-11}$~erg~cm$^{-2}$~s$^{-1}$. At the distance of UKS 1  \citep[$d=$~15.6~kpc][]{Baumgardt2021MNRAS},  
this translates to a $\gamma$-ray luminosity of 
$L^{\rm LAT}_{\gamma}=(9.44\pm7.97)\times10^{35}$~erg~s$^{-1}$. 
Based on 3HWC survey data, the differential flux at the peak position of the TeV feature is found to be $f^{\rm 3HWC}_{\gamma}=(3.27\pm1.61)\times10^{-14}$~cm$^{-2}$~s$^{-1}$~TeV$^{-1}$.   


\section{Physical Implications}
Assuming an average MSP spin-down power of $\langle\dot{E}\rangle \sim 2\times10^{34}\,{\rm erg\,s^{-1}}$ with a conversion efficiency $\eta_\gamma \sim 0.08$, the $\gamma-$ray luminosity $L^{\rm LAT}_{\gamma}$ of UKS~1 suggests it may host $\sim 100$ MSPs which can provide a strong collective pulsar wind outflow.

Given its high velocity of $\sim270$ km s$^{-1}$ moving around the Galactic disk \citep{2020A&A...643A.145F}, the GC wind is expected to be confined by a bow shock at $R_{bs} \sim 0.6$ pc, comparable to the cluster core radius. Relativistic leptons accelerated at this shock primarily cool through ICS. Due to UKS~1's relatively low stellar luminosity ($\sim4\times10^4 L_\odot$), the cooling length for multi-TeV particles reaches $\sim100$ pc, consistent with the observed offset of $0.39^\circ$ at 15.6 kpc.

Compared to the other TeV nebula around Terzan 5, UKS~1 exhibits a much larger TeV-to-GeV flux ratio, plausibly due to weaker starlight fields and, hence, less severe radiative losses \citep{HESS2011AnA,Terzan5_LAT}. This highlights UKS~1 as an unusual case among GCs, capable of sustaining extended, displaced TeV emission.

\section{Future Prospects}
GCs such as UKS~1 that move supersonically through the Galactic disk are natural laboratories for studying bow shock physics and cosmic ray transport on 10--100 pc scales. The supersonic motion through the disk not only confines the GC wind but may also seed turbulence that affects the scattering and diffusion of cosmic rays into the Galactic environment. 
Future observations with the Cherenkov Telescope Array Observatory (CTAO) can constrain cosmic ray scattering processes in GC magnetotails. Establishing GCs as a TeV emitter would have broader implications for understanding how MSP winds contribute to the Galactic cosmic ray population \citep{2018JKAS...51..171H}.

Recently, \citet{Owen_ICRC25} showed that GCs can be major contributors to the $\gamma$-ray emission of massive quiescent galaxies. The GeV-TeV spectrum from a GC is governed by several environmental parameters: luminosities of the stars and MSPs hosted by the cluster, the strength of the local soft photon field, and the ambient ISM density. Since these factors vary with galactocentric radius, the $\gamma$-ray properties of a galaxy’s GC system is expected to depend on the spatial distribution of its clusters. The distribution of GCs is closely related to the assembly history of a galaxy. Therefore, once we unveil how different galaxy formation scenarios imprint distinct spectral signatures on the $\gamma$-ray emission from GC systems, observations in GeV-TeV regime may be able to trace galactic evolutionary pathways.

\bibliographystyle{iaulike}
\bibliography{reference}

\begin{thebibliography}{}

\bibitem[{Abdo} et~al., 2010]{GC_LAT}
{Abdo}, A.~A., {Ackermann}, M., {Ajello}, \& {Fermi LAT Collaboration} 2010, {A
  population of gamma-ray emitting globular clusters seen with the Fermi Large
  Area Telescope}.
\newblock {\em Astronomy and Astrophysics}, 524, A75.

\bibitem[{Albert} et~al., 2020]{3hwc}
{Albert}, A., {Alfaro}, R., {Alvarez}, C., {Camacho}, \& {HAWC Collaboration}
  2020, {3HWC: The Third HAWC Catalog of Very-high-energy Gamma-Ray Sources}.
\newblock {\em Astrophysical Journal}, 905(1), 76.

\bibitem[{Ballet} et~al., 2023]{4fgldr4}
{Ballet}, J., {Bruel}, P., {Burnett}, T.~H., {Lott}, B., \& {The Fermi-LAT
  collaboration} 2023, {Fermi Large Area Telescope Fourth Source Catalog Data
  Release 4 (4FGL-DR4)}.
\newblock {\em arXiv e-prints},, arXiv:2307.12546.

\bibitem[{Baumgardt} and {Vasiliev}, 2021]{Baumgardt2021MNRAS}
{Baumgardt}, H. \& {Vasiliev}, E. 2021, {Accurate distances to Galactic
  globular clusters through a combination of Gaia EDR3, HST, and literature
  data}.
\newblock {\em Monthly Notices of the Royal Astronomical Society}, 505(4),
  5957--5977.

\bibitem[{Bednarek} and {Sitarek}, 2007]{Bednarek2007MNRAS}
{Bednarek}, W. \& {Sitarek}, J. 2007, {High-energy {\ensuremath{\gamma}}-rays
  from globular clusters}.
\newblock {\em Monthly Notices of the Royal Astronomical Society}, 377(2),
  920--930.

\bibitem[{Bednarek} and {Sobczak}, 2014]{Bednarek_2014}
{Bednarek}, W. \& {Sobczak}, T. 2014, {Misaligned TeV {\ensuremath{\gamma}}-ray
  sources in the vicinity of globular clusters}.
\newblock {\em Monthly Notices of the Royal Astronomical Society}, 445(3),
  2842--2847.

\bibitem[{Cheng} et~al., 2010]{Cheng_2010}
{Cheng}, K.~S., {Chernyshov}, D.~O., {Dogiel}, V.~A., {Hui}, C.~Y., \& {Kong},
  A.~K.~H. 2010, {The Origin of Gamma Rays from Globular Clusters}.
\newblock {\em Astrophysical Journal}, 723(2), 1219--1230.

\bibitem[{Fern{\'a}ndez-Trincado} et~al., 2020]{2020A&A...643A.145F}
{Fern{\'a}ndez-Trincado}, J.~G., {Minniti}, D., \& {et al.} 2020, {The
  enigmatic globular cluster UKS 1 obscured by the bulge: H-band discovery of
  nitrogen-enhanced stars}.
\newblock {\em Astronomy and Astrophysics}, 643, A145.

\bibitem[{H.~E.~S.~S. Collaboration et al.}, 2011]{HESS2011AnA}
{H.~E.~S.~S. Collaboration et al.} 2011, {Very-high-energy gamma-ray emission
  from the direction of the Galactic globular cluster Terzan 5}.
\newblock {\em Astronomy and Astrophysics}, 531, L18.

\bibitem[{Hui}, 2018]{2018JKAS...51..171H}
{Hui}, C.-Y. 2018, {A Golden Decade of Gamma-Ray Pulsar Astronomy}.
\newblock {\em Journal of Korean Astronomical Society}, 51(6), 171--183.

\bibitem[{Hui} et~al., 2010]{Hui2010.714}
{Hui}, C.~Y., {Cheng}, K.~S., \& {Taam}, R.~E. 2010, {Dynamical Formation of
  Millisecond Pulsars in Globular Clusters}.
\newblock {\em Astrophysical Journal}, 714(2), 1149--1154.

\bibitem[{Hui} et~al., 2011]{Hui_2011}
{Hui}, C.~Y., {Cheng}, K.~S., {Wang}, Y., {Tam}, P.~H.~T., {Kong}, A.~K.~H.,
  {Chernyshov}, D.~O., \& {Dogiel}, V.~A. 2011, {The Fundamental Plane of
  Gamma-ray Globular Clusters}.
\newblock {\em Astrophysical Journal}, 726(2), 100.

\bibitem[{Kong} et~al., 2010]{Terzan5_LAT}
{Kong}, A.~K.~H., {Hui}, C.~Y., \& {Cheng}, K.~S. 2010, {Fermi Discovery of
  Gamma-ray Emission from the Globular Cluster Terzan 5}.
\newblock {\em Astrophysical Journal Letters}, 712(1), L36--L39.

\bibitem[Krumholz et~al., 2024]{Krumholz_2024}
Krumholz, M., Crocker, R., Bahramian, A., \& Bordas, P. 2024, Teraelectronvolt
  gamma-ray emission near globular cluster terzan 5 as a probe of cosmic ray
  transport.
\newblock {\em Nature Astronomy}, 8, 1284--1293.

\bibitem[{Owen} et~al., 2025]{Owen_ICRC25}
{Owen}, E.~R., {Inoue}, Y., {Hui}, C.-Y., {Fujiwara}, T., \& {Kong}, A. K.~H.
  2025, {GeV-TeV Connections in Galaxies: Evolutionary Signatures from Pulsars
  in Globular Clusters}.
\newblock {\em arXiv e-prints},, arXiv:2508.16925.

\bibitem[{Shin} et~al., 2025]{Shin2025A&A}
{Shin}, J., {Hui}, C.~Y., {Kim}, S., {Oh}, K., \& {Owen}, E.~R. 2025, {A
  possible GeV-TeV connection in the direction of the globular cluster UKS 1}.
\newblock {\em Astronomy and Astrophysics}, 696, L11.

\bibitem[{Song} et~al., 2021]{Song_2021}
{Song}, D., {Macias}, O., {Horiuchi}, S., {Crocker}, R.~M., \& {Nataf}, D.~M.
  2021, {Evidence for a high-energy tail in the gamma-ray spectra of globular
  clusters}.
\newblock {\em Monthly Notices of the Royal Astronomical Society}, 507(4),
  5161--5176.

\bibitem[{Tam} et~al., 2016]{Tam2016JASS}
{Tam}, P.-H.~T., {Hui}, C.~Y., \& {Kong}, A. K.~H. 2016, {Gamma-ray Emission
  from Globular Clusters}.
\newblock {\em Journal of Astronomy and Space Sciences}, 33(1), 1--11.

\end{thebibliography}


\end{document}